# Earthquakes and related anomalous electromagnetic radiation

Title shortened form: Earthquakes electromagnetic anomalies

Manana Kachakhidze[1], Nino Kachakhidze-Murphy[1], Badri Khvitia[2]

[1]Georgian Technical University, Tbilisi, Georgia
[2]Sokhumi Institute of Physics and Technology

**Corresponding author** Kachakhidze M. email: kachakhidzem@gmail.com
Nino Kachakhidze-Murphy email: n.k.kachakhidze@gmail.com
Badri Khvitia  email:  badri.khvitia@gmail.com

**Abstract**

According to the presented work, VLF/LF electromagnetic emissions might be declared as the main precursor of earthquakes since based on these very emissions, it might predict ($M \geq 5$) inland earthquakes. As for ULF radiations, it governs some processes going on in the lithosphere-atmosphere-ionosphere coupling (LAIC) system. By these points, VLF/LF/ULF electromagnetic emissions have to consider more universal fields than other geophysical field anomalies during the earthquake preparation period up to aftershocks extinction.

## 1. Introduction

Nowadays, the problem concerning to investigation of the relationship between electromagnetic emissions and tectonic processes in the earth's crust is very important.
 This article is devoted to the study of these connections.
 Very significant papers are published in the scientific world on the basis of ground-based and satellite data of earth VLF/LF and ULF electromagnetic (EM) emissions observed in the earthquake preparation period (Molchanov, et al., 1998, 2008; Hayakawa, et al., 1996;2013; 2019; 2021; Biagi, 1999; Biagi et al., 2014;2019. Uyeda, et al., 2000;  Uyeda, 2013; Hattori, et al., 2004; Hattori, 2004;  Freund, 2000; Freund, et al., 2006; Parrot, 2006;  Pulinets, et al., 2006;  Pulinets, 2009; Eftaxias, et al., 2009; 2010, 2018).
 These phenomena are detectable both at the laboratory and geological scale (Molchanov, et al., 1998, 2008; Hayakawa, et al., 1996;2013; 2019;2021; Eftaxias, et al., 2009; 2010, 2018)
 Observations proved that when a material is strained, electromagnetic emissions in a wide frequency spectrum ranging from kHz to MHz are produced by opening cracks (Eftaxias, et al., 2009; 2010, 2018). On the large (geological) scale, intense MHz and kHz EM emissions precede earthquakes that: (i) occurred on land (or near the coastline),  (ii) were large (magnitude 6 or larger), or (iii) were shallow. (Eftaxias, et al., 2009; 2010, 2018; Karamanos, et al., 2006). Importantly, the MHz radiation precedes the kHz at geophysical scale, as well. These emissions are constantly accompanied by ULF radiations  (Eftaxias, et al., 2009; Kapiris, et al., 2004a;).
 Our goal is to consider the possible mechanisms for the origination of electromagnetic radiation before an earthquake.

## 2. Discussion

2.1. VLF/LF electromagnetic emission detected prior to the large earthquakes

Earth crust segment where incoming earthquake focus is to be formed from the very starting moment of earthquake preparation belongs to the system, which suffers specific type oscillations: the process of energy accumulation is in progress in the system. But simultaneously, as a result of foreshocks, main shock, and aftershocks, the accumulated energy is released too. With this, in view, this system is an oscillation system.

The extreme diversity of oscillation systems and their properties needs the identification of common features in various oscillation systems and their gathering into certain classes and types according to the most characteristic signs.

In the seismogenic area, the mass, elasticity (mechanical systems), capacity, and inductance (electric systems) elements are uniformly and uninterruptedly spread in the whole volume of the system (Migulin et al., 1978). Moreover, in the earthquake preparation area, each least element has its own capacity and inductance because of piezo-electric, piezo-magnetic, electrochemical, and other effects. Therefore seismogenic zone simultaneously can be considered a distributed system. The seismogenic area can be determined by the Dobrovolsky formula (1) (Dobrovolsky et al. (1979)

$$R = 10^{0.43M} \qquad (1)$$

where $R$ is the strain area radius in kilometer and $M$ is earthquake magnitude.

Since in this system, permanently acts the tectonic stress that changes the physical and chemical properties of the environment, the relationship between the locations of elements (or groups of elements) plays a significant role by point of view of the functioning of the system

Usually, in distributed systems, it is impossible to isolate a single point, since each channel for the passage of energy is considered a pair of poles (connection points) (Migulin et al., 1978), therefore, the radiated in the system frequency will change, in accordance with changes in the lengths of the combined fractures involved in the process of the formation of the fault.

Just on these considerations was based the model of the generation of electromagnetic radiation existed prior to the earthquake, where the formula for the length $l$ of the earthquake fault is obtained:

$$l = k \frac{c}{\omega} \qquad (2)$$

where $\omega$ is the eigen-frequency of electromagnetic emissions, $c$ is the light speed, $k$ is the characteristic coefficient of geological medium (it approximately equals to 1).

In the scientific literature, electromagnetic radiation before an earthquake has recently been grouped as follows:

ULF (ultra-low frequency, **f<1Hz**); ULF, ELF (extremely low frequency, **1Hz<f<3kHz**); VLF (very low frequency, **3kHz<f<30kHz**); LF (low frequency, **30kHz<f<300 kHz**) (Hayakawa et al., 2019). The question arises whether this radiation is earthbound or not during the earthquake preparation period.

To clear this issue, let's separately consider the frequencies of the range of electromagnetic radiation mentioned above.

According to the formula (2), the frequencies of the spectrum in the range of 23 830 kHz≥f≥0.378 kHz, corresponding to the lengths of the faults of the 1≤M≤9 magnitudes earthquakes, originate in the earthquake preparation process.

- ULF (ultra-low frequency, **f<1Hz**) - is outside this range;
- ULF, ELF (extremely low frequency, 1Hz<f<3kHz) is partially included in this range for 9.0 ≥M ≥7.5 magnitude earthquakes, the corresponding frequency range of which is 0.4 kHz < f < 3kHz.
- VLF (very low frequency, **3kHz<f<30kHz**). Electromagnetic radiation of this range corresponds to earthquakes of magnitude 7.5≥M ≥5.8;
- LF (low frequency, **30kHz<f<300 kHz**). Electromagnetic radiation of this range corresponds to earthquakes of magnitude 5. 8≥M ≥4.1

Thus, the source of LF, VLF and ELF (0.4 kHz<f<3kHz) frequencies during the earthquake preparation period is the earth (Kachakhidze et. al, 2015,2019, 2022). Electromagnetic radiation of high frequencies almost does not reach the earth's surface.

We focus on electromagnetic radiation in the 102 kHz – 0.377 kHz frequency range since earthquakes of magnitude 5≤M≤9 are noteworthy for seismically active regions and countries.

The main determining parameter of the magnitude of an earthquake is the length of the fault formed in the earthquake focus.

If any geophysical field, in addition to determining the epicenter and time of occurrence of the earthquake, can analytically describe the change in the length of the fault in the focus during the preparation of the earthquake, it is clear that only such a field can be considered as a precursor of the earthquake.

The other fields, which change abnormally during the earthquake preparation period, but cannot describe the change in the length of the fault, should be considered only as indicators.

Since it was found that the electromagnetic wave of VLF/LF frequency before the earthquake is formed by the complex geological process of the coalescence of cracks in the earthquake focus, and therefore, the frequency ω corresponding to this wave is the parameter that analytically describes the changes of the fault length arisen in the focus during the earthquake preparation period, the electromagnetic radiation of the VLF/LF frequency should be considered as a precursor to an earthquake.

In addition to the electromagnetic radiation of VLF/LF frequencies, we are also interested in the anomalous perturbation of ULF electromagnetic radiation, which constantly accompanies the process of earthquake preparation.

2.2. ULF electromagnetic emission detected prior to the large earthquakes.

It is proved, that dynamic processes in the earthquake preparation zones can produce current systems of different kinds (Molchanov et al.,1998; Kopytenko, et al., 2001) which can be local sources for electromagnetic waves at different frequencies, including ULF. ULF waves can propagate through the crust and reach the earth's surface, unlike high-frequency waves (Hattori, et al., 2004; Kopytenko, et al., 2006; Liu, et al., 2006; Varotsos, et al., 2011; Chen, et al., 2011).

Thus, in ground-based observations, we could expect some ULF signals of seismic origin observed in both geo-electric and geomagnetic fields (Kopytenko, et al., 2001; Uyeda, 2013).

Different attempts have been made to explain the generation of electro-telluric variations before an earthquake takes place in an earthquake preparation zone. Their disturbances in the preparation zone were considered a factor of such importance that in a number of works it was proposed to use telluric variations as a short-term precursor of strong earthquakes (Varotsos et.al., 2006, 2011; Lazarus, 1993;).

We have a different view on this issue.

As a result of the growth of tectonic stress heterogeneity appears in earthquake preparation areas, but earthquake preparation takes place in a relatively weak zone, by the view of solidity (Morozova, et al., 1999; Tada-nori Goto, et al., 2005; Kovtun, et al., 2009, Freund, 2000).
Since tectonic stress basically "works" only for the formation of the main fault in the earthquake focus, in other parts of the seismogenic area, it cannot cause the fracturing of rocks and their significant combining, that is, it can no longer form the necessary conditions for the occurrence of an earthquake.However, tectonic stress can cause perturbations of the geophysical fields in the seismogenic area.

One such field is the telluric current.

As a result of the growth of tectonic stress heterogeneity appears in earthquake preparation areas(Freund, 2000), similar to "Frankel's generator" this segment of the earth's crust will have inductive polarization (Frenkel, 1949; Yoshino, 1991; Molchanov et al., 1998; Liperovsky et al., 2008).

Generally, polarization charge should be distributed over some surface, which should be limited by fault or should be formed along the faults (Yoshino, 1991).

Experimentally an important fact has been proved that at the formation of cracks in the earthquake preparation period, electric dipoles appear on their surface (Freund, et al. 2006; Eftaxias, et al.,2009, 2010, Eftaxias, et al., 2018).

The polarization charge takes part in two different processes:

**a)** If anywhere in the seismogenic area, microcracks coalesce in the form of any size rupture nucleus (including the main fault) and in the fault plane there are changes in the specific electrical resistance of the rocks, that is there are inclusions of high electric conductivity, which conditions the sharp increase of the electric conductivity, it is not excluded that the layer on which the polarization charge is distributed and the fault, like a double-wire conduction layer might be locked by vertical electric field and form an oscillating contour-like structure. This will happen if the segment of the earth's crust, on the surface of which the polarized charges are distributed, is about the size of the fault and is formed approximately along this fault. When the tectonic stress overcomes the limit of the geological strength of the medium it begins the rock integrity fracturing process (later it goes to fault formation avalanche process and ends with an earthquake) which accompanies by emissions of electromagnetic waves of VLF/LF frequency. The value at this time emitted frequency of the electromagnetic wave depends on the length of the fault.

In the case of small cracks, the electromagnetic radiation will be of the order of MHz, and in the case of cracks of the order of *km,* it goes into kHz (Kachakhidze, et al., 2015, 2019, 2022).

**b)** In areas where tectonic stress is not able to form a crack, and because, in general, rock density increases with depth, in case of the same tectonic stress effect, more inhomogeneities appear in the upper rocks with lower density compared to the lower rocks, that is, in the upper rocks, more polarization charges are generated compared to the lower ones.

However, as stated above, since the polarization charge must be distributed over some surface (Yoshino, 1991), it is clear that each of these surfaces will be approximately equipotential.

In general, the work done during the movement of the charge $q_0$ on the $\Delta l$ element of the equipotential surface is equal Vepkhvadze, 1995):
$$\Delta A = q_0(\varphi_1 - \varphi_2) \quad (3)$$
This work can also be represented by means of field tension:
$$\Delta A = q_0 E \Delta l \cos\alpha \quad (4)$$
According to formulas (3) and (4):

$$q_0 E \Delta l \cos\alpha = 0$$
and since $q \neq 0, E \neq 0, \Delta l \neq 0$, therefore $\cos\alpha = 0$ i.e. $\alpha = \frac{\pi}{2}$.

Since the field lines in any electrostatic field are perpendicular to the equipotential surfaces and in addition in the case of the field of the system of charges, the tensions summarize vectorially, it is obvious that the total direction of the electric field tensions created by the polarization charges of these inhomogeneities will be the same.

It is known that the telluric electric fields all the time change by direction and magnitude at any point (Kraev, 2007), At the same time at perturbation telluric field becomes linearly (or plainly) polarized (Kraev, 2007).

Therefore, under conditions of increasing tectonic stress, we have to assume that the telluric field is mainly disturbed and polarized.

If we exclude the effects that external factors can cause, the changes in this field in a given area will uniquely depend on the changes in tectonic stresses.

In generaly, in special scientific works magneto-telluric field is considered as a field that is perturbed by local and regional factors (Kraev, 2007).

The telluric field perturbation and polarization during earthquake preparation are contributed by: a) stratospheric-electrical processes (ionospheric oscillations, auroras), b) boundary-electrical processes (filtration-electrical processes, convection currents in the lower layers of the atmosphere, lightning processes, etc.), c) lithospheric-electrical processes (contact voltages, thermoelectric and chemical-electrical processes (Kraev, 2007);

It is known that thanks to contact of solid or gaseous phases existing between two mediums, the earth and atmosphere, diffusion of electrons and ions and ion adsorption take place, which conditions the creation of a stable electric layer (dipole layer) on the contact. In this layer, the electric field, supported by factors conditioned by earthquake preparation, can be called an "additional" electric field and can be marked as $E_n$ (5) (Kraev, 2007) (Fig, 1).

In this case, as has been mentioned above, electric field potential at the separating border of these two mediums suffers discontinuation, which equals contact "additional" electric field strength:

$$\varepsilon^{add} = \int_1^2 E_n^{add} dn \quad (5)$$

where 1 and 2 points are located on both near coasts of the contact surface. It is clear that the fact of mentioned field discontinuation will be expressed in all geophysical phenomena connected with the "additional" field.

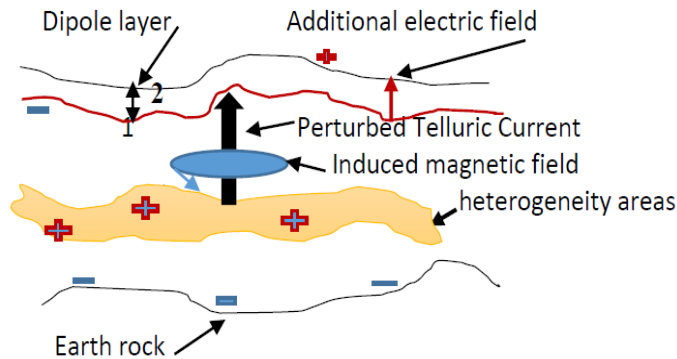

Fig. 1. Perturbed Telluric current

Since the charges move in the direction of the field line, it is obvious that at this time is generated, the stress-induced electric current in the rock caused by the changes in mobility of charged dislocations (Tzanis and Vallianatos, 2002; St-Laurent et al., 2006; Triantis, et al., 2008) and/or point defects (Freund, 2000).

Telluric current is practically a synonym for geoelectric potential difference and preseismic telluric current signals, called seismic electric signals (SES) (Orihara et al., 2012),
Obviously, the direction of the telluric current, under these conditions, will be polarized and directed vertically in the direction of the Earth-ionosphere. This fact can also be observed in nature (Pulinets, 2009).

As this current is associated with the displacement of charges, it releases heat (changing the orientation of the dipoles causes heat release) (Vepkhvadze, 1995), so the temperature in this area will increase, which is also confirmed by the experiment (Tramutoli, et al., 2005; Pulinets, et al., 2006; Saradjian, et al., 2011; Kundu et al., 2022).

The telluric current will generate a magnetic field by induction, which is known in the scientific literature too: as a result of the current changes in the Earth's crust the anomalous magnetic field variations start (which, according to Maxwell equations, should be accompanied by a strong SES activity) (Chapman, S. and Whitehead, 1922; Panayiotis et al., 2019).

Indeed, according to Maxwell's theory, in the case of plane electromagnetic waves, a change in the electric field directed along the axis OZ leads to the generation of a magnetic field directed along the axis OY.

Since in our case, the telluric current is directed from the Earth to the ionosphere along OZ, and it is also flatly (or linearly) polarized, for the magnetic field induced by this field we have:

$$\varepsilon\varepsilon_o \ \frac{\partial E_z}{\partial t} = \frac{\partial H_y}{\partial x} \quad (6)$$

i.e. the induced magnetic field will also be polarized due to the polarization of the OY component (Fig. 1). These magnetic field anomalies can also be observed in nature during earthquakes (Liu, et al, 2006; Bleier, et al., 2009, 2010; Chen C. H.,, et al, 2013; Chen, H et al., 2022; Hayakawa et al., 1996, 2019, 2021)

The electric field at the separating border of two mediums - the earth, and atmosphere - suffers discontinuation, but the induced magnetic field spreads in the zone between the earth and the ionosphere, creating a polarized eddy electric field in the atmosphere.

When perturbed telluric current causes atmospheric electric field polarization, the principle of global electric contour closure is destroyed, and in the global electric circuit system a defined anomalous zone appears. In this case lines of force of global atmospheric electric current are no more closed, that is, for a definite time, they do not go beyond the earth-ionosphere boundary and accumulate charges only at the ends of lines of force.

The perturbed magneto-telluric field should be revealed itself on the very first stage of formation of the main fault; this fact was confirmed by laboratory and field observations (Molchanov et al., 2008; Orihara, et al., 2012; Panayiotis et al., 2019;).

In addition, experimental investigations proved that the earth's currents generate instant magnetic variations (Kraev, 2007)). Besides during earthquake preparation in the epicentral area frequencies of electric and magnetic fields should be equal, which was proved experimentally (Hattori, 2004; Hattori et. al., 2004).

It should be taken into consideration that magneto-telluric field perturbation will take place not only during the period preceding the earthquake but also after it too, till tectonic stress accumulated in the focal area is released completely (Moroz, et al., 2004).

Thus, during the preparation, occurrence, and subsequent period of the earthquake, including the aftershocks attenuation, changes in the telluric current are directly related to changes in tectonic stress, so this field to be considered only as an indicator of an earthquake.

## 3. Summary

Accumulation of tectonic stress in any area of a seismically active region (country) causes:
On the one hand:
- Formation of a fault in the focus of the incoming earthquake.
- This process is accompanied by the generation of VLF/LF electromagnetic waves, based on the data of which it is possible to determine simultaneously the magnitude, location, and time of the expected earthquake;

On the other hand:
- In the area of accumulation of tectonic stress, a vertical telluric current of the earth-ionosphere direction is generated;
- The polarized telluric current in the earth-atmosphere border suffers discontinuation;
- Polarized telluric current, due to the polarization of the eastern component of the Earth's magnetic field, causes the generation of a polarized magnetic field in the earthquake preparation area;
- -The polarized magnetic field uninterruptedly passes through the earth-atmosphere border;
- The earth's surface should have positive potential permanently in the earthquake preparation area for a rather long period.

Thus:
- Tectonic stress causes and fully governs ULF electromagnetic field anomalies in the earthquake preparation area, which from its side conditions continuous interconnection of the lithosphere-atmosphere-ionosphere (LAIC) system.
- ULF electromagnetic radiation is only an indicator of an earthquake since anomalous changes of this field are merely related to changes in tectonic stress, and not to the process of fault formation in the focus of the incoming earthquake.